\documentclass[twoside,showpacs,reprint,superscriptaddress,amsmath,amssymb,aps,floatfix]{revtex4-1}
\usepackage[utf8]{inputenc}
\usepackage{amsmath}
\usepackage{amssymb}
\usepackage{graphicx}
\usepackage{breqn}
\usepackage{mathtools}
\usepackage{amsthm}
\usepackage{color}
\usepackage{soul}
\usepackage{breqn}
\usepackage{tikz}
\usepackage{etoolbox}
\usepackage{svg}
\usepackage[nodisplayskipstretch]{setspace}
\usepackage{float}
\usepackage{epstopdf}
\usepackage[justification=centering]{caption}
\usepackage{cancel}

\makeatletter
\preto\maketitle{%
  \begingroup\lccode`~=`,
  \lowercase{\endgroup
  \let\saved@breqn@active@comma~% save breqn active comma
  \let~}\active@comma % set the active comma to what revtex4-1 wants
}
\appto\maketitle{%
  \begingroup\lccode`~=`,
  \lowercase{\endgroup
  \let~}\saved@breqn@active@comma % undo the change
}

\usepackage{bm}% bold math
\makeatother

\def\be{\begin{eqnarray}}
\def\ee{\end{eqnarray}}
\def\r{{\bf r}}

\def\E{{\bf E}}
\def\H{{\bf H}}

\def\e{\bm{\hat{e}}}

\def\im{{\rm i}}

\begin{document}
\title{Enhanced spin-orbit optical mirages from dual nanospheres}

\author{Jorge Olmos-Trigo}
\affiliation{Donostia International Physics Center (DIPC),  20018 Donostia-San Sebasti\'{a}n,  Spain}

\author{Cristina Sanz-Fern\'andez}
\affiliation{Centro de F\'{i}sica de Materiales (CFM-MPC), Centro Mixto CSIC-UPV/EHU,  20018 Donostia-San Sebasti\'{a}n,  Spain} 

\author{Aitzol Garc\'{i}a-Etxarri}
\affiliation{Donostia International Physics Center (DIPC),  20018 Donostia-San Sebasti\'{a}n,  Spain}
\affiliation{Centro de F\'{i}sica de Materiales (CFM-MPC), Centro Mixto CSIC-UPV/EHU,  20018 Donostia-San Sebasti\'{a}n,  Spain}

\author{Gabriel Molina-Terriza}
\affiliation{Donostia International Physics Center (DIPC),  20018 Donostia-San Sebasti\'{a}n,  Spain}
\affiliation{Centro de F\'{i}sica de Materiales (CFM-MPC), Centro Mixto CSIC-UPV/EHU,  20018 Donostia-San Sebasti\'{a}n,  Spain}
\affiliation{IKERBASQUE, Basque Foundation for Science, 48013 Bilbao, Spain}

\author{F. Sebasti\'an Bergeret}
\affiliation{Donostia International Physics Center (DIPC),  20018 Donostia-San Sebasti\'{a}n,  Spain}
\affiliation{Centro de F\'{i}sica de Materiales (CFM-MPC), Centro Mixto CSIC-UPV/EHU,  20018 Donostia-San Sebasti\'{a}n,  Spain}

\author{Juan Jos\'e S\'aenz}
\email{juanjo.saenz@dipc.org}
\affiliation{Donostia International Physics Center (DIPC),  20018 Donostia-San Sebasti\'{a}n, Spain}
\affiliation{IKERBASQUE, Basque Foundation for Science, 48013 Bilbao,  Spain}

\begin{abstract}
%It is well-known that the spiral structures induced by the Poynting vector lead to a subwavelength displacement of an electric dipole position. This apparent shift does not depend on the optical properties but on the scattering angle, reaching its maximum at the perpendicular direction to the incoming wave. However, as we will show, the interaction between the electric and magnetic dipoles leads to an optical mirage that depends strongly on the optical target response. This interference effect gives rise to a considerably enhancement (far above the wavelength)  of the optical mirage close to backscattering, when the dual symmetry is satisfied (first Kerker condition).
{Spin-orbit interaction of light} can lead to the so-called optical mirages, i.e. a perceived displacement {in} the position of {a} particle due to the spiraling structure of the scattered light. In electric dipoles, the maximum displacement is subwavelength and does not depend on the optical properties of the scatterer.   Here we will show that the optical mirage {in}  high refractive index dielectric nanoparticles depends {strongly} on the ratio between electric and magnetic dipolar responses. When the dual symmetry is satisfied (at the first Kerker condition), there is a considerable enhancement (far above the wavelength) of the spin-orbit optical mirage which can be related to the emergence of an optical vortex in the backscattering direction.
\end{abstract}

\maketitle
      
\setstretch{1}
%In addition to energy and linear momentum a light wave  carries angular momentum  \cite {allen2003optical}.
%\FSB{The later consists of a spin (\emph{SAM}) and orbital angular momentum (\emph{OAM}). The study of the  possible momentum transfer between these two contributions, {\it i.e.} the  spin-orbit interaction (SOI), has attracted a great deal of interest in the past years\cite{berry2005orbital,bliokh2015spin}. }.  

It is customary to separate the angular momentum (AM) of light \cite{allen2003optical} into two contributions, the spin (SAM) and the orbital angular momentum (OAM), that can be coupled by   light propagation and scattering.    The study of this spin-orbit interaction (SOI) has attracted a great deal of interest in the past years \cite{liberman1992spin,crichton2000measurable,berry2005orbital,bliokh2015spin}.

An interesting  analogy between the SOI in light and  the spin Hall effect  (SHE) in  electronic systems can  be drawn\cite{dyakonov1971current,hirsch1999spin}.  In the latter, electrons with different spins  are deflected differently by scattering off impurities due to the SOI. This leads to a transversal spin current that in turn induces a measurable spin accumulation at the sample edges. One of the microscopic origins of the SHE is the so-called side-jump mechanism \cite{berger1970side}, in which a spin-dependent displacement of the center of mass of the electronic  wave packet takes place due to the SOI (for more details we refer to the  reviews  \cite{dyakonov2008spin,sinova2015spin}).

Similarly, an apparent transversal displacement of a target particle  induced by light scattering can be explained by an AM exchange.    Hereafter this effect is  referred to as  \emph{optical mirage} and  has been observed in several situations, for example, in  beams impinging on a dielectric surface \cite{onoda2004hall,bliokh2006conservation,hosten2008observation} or when considering a  spherical target  described by a single electric polarizability \cite{haefner2009spin}.  
In the latter case, the apparent shift of the dipole localization does not depend on the optical properties,  but on the scattering angle, with opposite displacements for incident left and right circularly polarized photons (spins). The apparent shift ($\Delta$) is maximized   at the {plane} perpendicular to the  direction of the incoming wave taking a value of $\Delta =\lambda/\pi$ and thus, it is always subwavelength.

 In this Letter, we demonstrate that by taking into account both, the electric and magnetic dipoles sustained by a high refractive index spherical particle, the subwavelength maximum limit can be drastically surpassed when the particle is excited by circularly polarized light. In other words, 
 a large  macroscopic apparent shift ($\Delta \gg \lambda$) is induced in the back scattering region.
 Specifically, we show that this optical mirage is related to the generation of a spiraling power flow and  can be explained in terms of an angular momentum redistribution per photon between the SAM and OAM contributions. Based on helicity conservation we predict an intriguing enhancement of the momentum transfer when the system is dual, i.e. when the electric and magnetic dipolar moments are equal. At this, so-called, ``first Kerker condition'' \cite{kerker1983electromagnetic,nieto2011angle,gomez2011electric}, the emitted light intensity vanishes in the backscattering direction, leading to the appearance of a ($2 \sigma$ charge) topological optical vortex.  
%\CSF{\st{This remarkable phenomena predicted and observed experimentally has attracted a large interest in the last few years. As we will show, at the first Kerker condition, there is a huge  enhancement of the apparent shift near the backscattering direction that is related to the appearance of an optical vortex in the reflected light.}}
\begin{figure}
\includegraphics[width=0.9 \columnwidth]{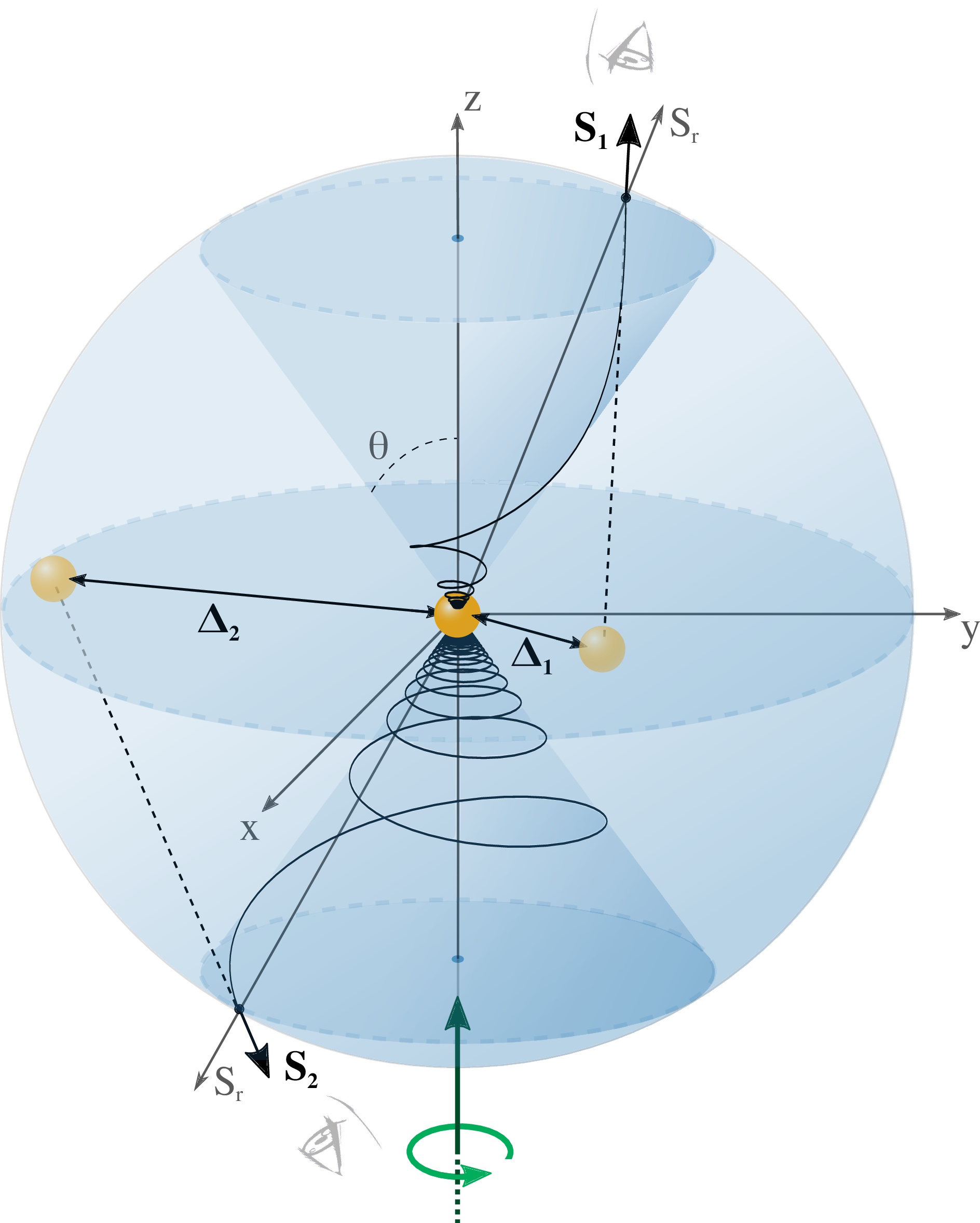}
\captionsetup{justification= raggedright}
\caption{Schematic representation of the optical mirage vector when considering a clockwise circularly polarized incoming wave (green straight  arrow lying on the $z$-axis). The observer, represented by an eye, perceives a non-radial scattered Poynting vector $({\bf{S}}_1,{\bf{S}}_2)$ that  leads to an apparent shift $(\bm{\Delta}_1,\bm{\Delta}_2)$ of the dipole localization, both lying  on the $xy$-plane. }\label{delta} 
\end{figure}
% \fixme{el texto que sigue "However,..." necesita ser re-escrito para dejar bien claro qué difrencia hay entre los nuestro y lo de los trabajos \cite{arnoldus2008subwavelength,haefner2009spin}. Intenta evitar los paréntesis en las  frases siguientes}\FSB{Yo pondría algo como, "However up to now the blabla has not been taken into account...." y luego abriría el párrafo siguiente con: In this letter, we tackle this problem and demonstrate....." y combinaria los dos párrafos siguientes en uno solo}
% However, as we will show when taking into account the interaction between the electric and magnetic dipoles (dipolar approximation), a considerably enhancement  (far above the wavelength) of this apparent shift (now optical mirage) can be obtained. 
%  We demonstrate that for a wave in a pure state of polarization, the spin-orbit interaction results in a spiraling power flow that is determined by the extent of the interaction. 
%\AGE{In particular}, we  consider a high refractive index spherical particle with strong magnetic and electric dipolar responses in the infrared as the target . % The asymmetry factor, $g = \langle \cos \theta \rangle$ \cite{gomez2012negative},  plays a crucial role in the optical response, leading to a macroscopic optical mirage value close to backscattering when the first Kerker condition  is satisfied \cite{kerker2016scattering, gomez2012negative}. 

We consider a non-absorbing dielectric sphere of radius $a$ and refractive index $n_{\rm{p}}$ embedded in an otherwise homogeneous  medium  with constant and real  refractive index $ n_{\rm{h}}$. 
The geometry of the scattering problem is sketched in Fig. \ref{delta}, where we consider a  circularly polarized plane wave with wavenumber  $k = n_{\rm{h}} k_0 = n_h 2 \pi / \lambda_0$ (being $\lambda_0$  the light wavelength in vacuum)
and helicity $
\sigma = \pm1$ (we associate left polarized light with a positive helicity  $\sigma = +1$) incident along the $z$-axis. 
%%%% AGE{
Instead of using the traditional multipole Mie expansion to describe the light scattered by the sphere \cite{jackson1999electrodynamics,bohren2008absorption},  we shall find it useful to work in a basis of multipoles, eigenfunctions, $\bm{\Psi}^{\sigma}_{lm}$,  of the helicity operator $\Lambda$, \cite{fernandez2012helicity,zambrana2013duality},
\be \bm{\Lambda} \bm{\Psi}^{\sigma}_{lm}= (1/k)\bm{\nabla} \times \bm{\Psi}^{\sigma}_{lm} = \sigma \bm{\Psi}^{\sigma}_{lm}, \nonumber 
\ee 
with 
\be
 \bm{\Psi}^{\sigma}_{lm}  &=& \frac{1}{\sqrt{2}}\left[\frac{\bm{\nabla} \times g_l(kr) {\bf{X}}_{lm} }{k} + \sigma g_l(kr) {\bf{X}}_{lm}\right], 
 \label{defPsi} \\
g_l(kr) &=& A_l^{(1)} h^{(1)}_l(kr) + A_l^{(2)} h^{(2)}_l(kr),\\
{\bf{X}}_{lm} &=& \frac{1}{\sqrt{l(l+1)}} {\bf{L}} Y_l^m(\theta,\varphi) ,
\ee
where, following Jackson's notation  \cite{jackson1999electrodynamics}, ${\bf{X}}_{lm} $ denote the vector spherical harmonic, with ${\bf{X}}_{00} = 0$, $g_l(kr)$ is a linear combination of the spherical Hankel functions, $Y_l^m(\theta,\varphi)$ are the spherical harmonics and $\bf{L}$ is the orbital angular momentum operator, ${\bf{L}} = -\im \ ( \r \times \bm{\nabla})$. 
In this helicity basis, the incident field can be written as
\be
 \frac{ \E^{(0)}_{\sigma}}{E_0} &=& \frac{\bm{\hat{x}} + \sigma \im \bm{\hat{y}}}{\sqrt{2}}  e^{\im kz} = \sum_{l=0}^\infty \sum_{m=-l}^{+l} \sum_{\sigma' = \pm1} C_{lm}^{\sigma \sigma'}    \bm{\Psi}^{ \sigma'}_{lm}, 
\\
 k Z\H^{(0)}_\sigma &=& -\im \bm{\nabla} \times \E^{(0)}_\sigma,
 \\ C_{lm}^{\sigma \sigma'} 
&=& \sigma \im^l \sqrt{8\pi(2l+1)} \delta_{m \sigma} \delta_{\sigma\sigma'},
 \ee
where $1/Z=\epsilon_0 c n_h$  (being $\epsilon_0$ and $c$  the vacuum permittivity and  speed of light, respectively) and
$ \bm{\Psi}^{ \sigma'}_{lm}$ is given by Eq. \eqref{defPsi} 
with $g_l(kr) = j_l(kr)$. 
 Such circularly polarized wave, with helicity $\sigma$, carries a $j_z = m=\sigma$ unit of total angular momentum per photon parallel to the propagation direction \cite{jackson1999electrodynamics}.  
 
 In the same basis,  the scattered fields are given by
\be
 \frac{\E^{\text{scat}}_\sigma}{E_0} &=&  \sum_{l=0}^\infty \sum_{m=-l}^{+l}  \sum_{\sigma' = \pm1} D_{l m}^{\sigma \sigma'}  \bm{\Psi}^{ \sigma'}_{lm} ,
\\ D_{lm}^{\sigma \sigma'}
&=& - \im^l \sqrt{4\pi(2l+1)}  \frac{\sigma a_l + \sigma'  b_l}{2} \delta_{m\sigma},
 \ee
where now, since they are outgoing waves at infinity, $g_l(kr) = h^{(1)}_l(kr)$.  Notice that  $a_l$, $b_l$ are the standard Mie electric and magnetic scattering coefficients \cite{bohren2008absorption}. Since a sphere presents axial symmetry around the $z$-axis, {\em the $j_z$ of the incident beam is preserved} and the scattered wave can only involve $m=\sigma$. {Consequently,} $\E^{\text{scat}}_\sigma$ is an eigenfunction of the $z$-component of the total (dimensionless) angular momentum operator, $\bf{J} = \bf{L}+\bf{S}^{\text{spin}}$ (as well as of ${\bf{J}}^{2}$)\cite{tischler2012role}, with eigenvalue $j_z=m=\sigma$,
  \begin{align}
 \sigma &=
\frac{{\E^{\text{scat}}_\sigma}^* \cdot \left( L_z    + {\bf{S}}^{\text{spin}}_z \right) \E^{\text{scat}}_\sigma}{\left|\E^{\text{scat}}_\sigma\right|^2}   = \ell_z(\r) + s_z(\r),
  \label{Jz}
   \\
  s_z(\r)  &= \frac{-\im  \left\{{\E^{\text{scat}}_\sigma}^* \times  \E^{\text{scat}}_\sigma\right\} \cdot \bm{\hat{e}}_z }{\left|\E^{\text{scat}}_\sigma\right|^2} 
  \\
  \ell_z(\r)  &= \frac{{\E^{\text{scat}}_\sigma}^* \cdot  L_z  \E^{\text{scat}}_\sigma}{\left|\E^{\text{scat}}_\sigma\right|^2}  =
    \frac{ -\im}{\left|\E^{\text{scat}}_\sigma\right|^2} 
 \left\{{\E^{\text{scat}}_\sigma}^* \cdot \frac{\partial \E^{\text{scat}}_\sigma}{\partial \varphi} \right\} \label{OAMpf}
 \end{align} 
Equation \eqref{Jz}  shows that the sum of the (dimensionless) {OAM}, $\ell_z(\r)$, and {SAM}, $s_z(\r)$, {\em per photon} is constant and equal to the helicity of the incoming plane wave.  Notice that this is valid even in the near field region and it would be valid even in the presence of absorption. 
However,  in general, the helicity is not preserved in the scattering process.
%The scattering of an electromagnetic wave by a sphere is a classical text-book problem that can be solved by expanding the fields in multipoles, which can be easily    Instead of using the well known 
% multipole expansion (see Chapter 10 in Ref. \cite{jackson1999electrodynamics}), 
% Let us define
%\be
%\xim_{\sigma}&=& \frac{ \bm{\hat{x}}  + \im \sigma \bm{\hat{y}}}{\sqrt{2}}
%= \frac{e^{\im \sigma \varphi}}{\sqrt{2}} \left( \sin\theta \e_r +
%\cos\theta \bm{\hat{e}}_\theta
%+ \im \sigma \bm{\hat{e}}_\varphi  \right) \CSF{Verified} \label{xivec}
% \\ \xim_0 &=& \bm{\hat{z}} = \cos\theta \e_r-\sin\theta \bm{\hat{e}}_\theta  \CSF{Verified}
%\ee
%the incident electric and magnetic fields can be expressed as
%\be \label{S:1}
%\E^{(0)} =  E_0e^{\im kz}  \xim_\sigma,&& \quad  Z \H^{(0)} = \im  \sigma  E_0 e^{\im kz} \xim_\sigma \CSF{Verified} \ee
%%%%%%AGE}
\begin{figure}
\includegraphics[width=0.9\columnwidth]{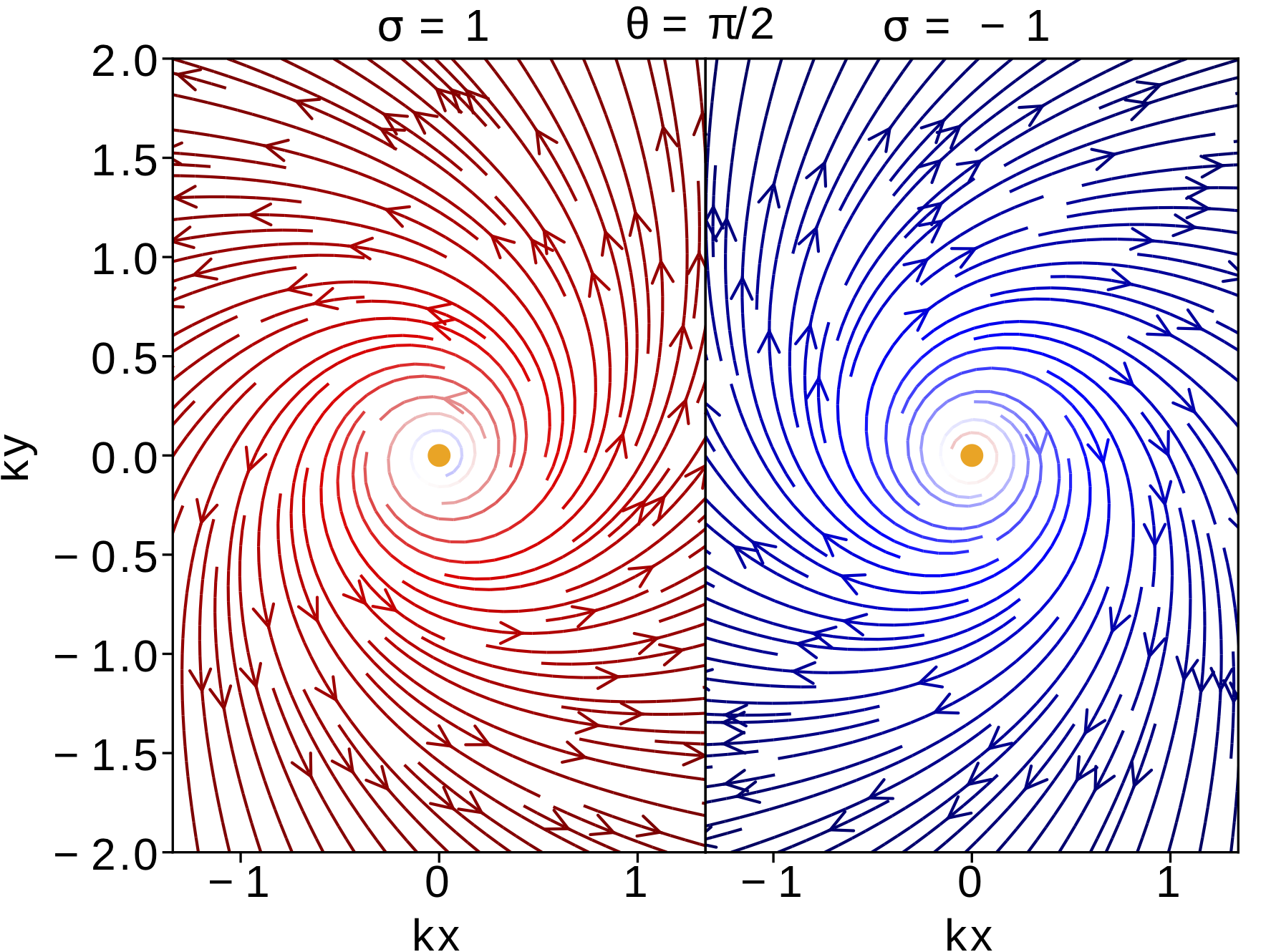}
\captionsetup{justification= raggedright}
\caption{Poynting vector streamlines with counterclockwise (clockwise) rotation  for  $\sigma =  1$ $(\sigma =-1)$ when viewed from the perpendicular direction, $\theta = \pi / 2$. This figure is valid for any dipolar response, i.e. arbitrary $\alpha_{\rm{E}}$ and  $\alpha_{\rm{M}}$. The orange circle represents the dipolar particle.}\label{flechas} 
\end{figure}

Let us now consider the scattering from a high refractive index (HRI) subwavelength sphere in a spectral range such that the optical response  can be described by  its first dipolar Mie coefficients $a_1$ and $b_1$, i.e. by its electric and magnetic polarizabilities
$\alpha_{\rm{E}}=\im a_1 (6\pi/k^3) $ and $\alpha_{\rm{M}}={\im}b_1(6\pi/k^3) $. 
%For relatively long wave-lengths, a lossless high refractive index (HRI)  particle with $n_p/n_h \gtrsim 2$ (and  $(n_p/n_h) 2\pi a/\lambda_0 \lesssim 4$)  can be characterized by only its  \cite{garcia2011strong,gomez2011electric}  (e.g. Silicon spheres of radius $a=230$nm in the telecom spectral range), where $a_1$ and $b_1$ are the first dipolar coefficients in the ``Mie'' multipolar expansion. 
The scattered field can  be written as the sum of two components with opposite helicity,
\be
 \frac{\E_{\sigma}^{\text{scat}}  }{E_0} &=&
  -   \frac{k^3}{\sqrt{6\pi}}  \Big\{ (\sigma \alpha_{\rm{E}} + \alpha_{\rm{M}}) \bm{\Psi}^{ +}_{1\sigma}
  +
   (\sigma \alpha_{\rm{E}} - \alpha_{\rm{M}}) \bm{\Psi}^{-}_{1\sigma} \Big\} \nonumber \\
 &=&  \E_{\sigma+} +  \E_{\sigma-}, \label{Eff}
\ee
which in far field limit become
\be
 \E_{\sigma \sigma'} &
  \sim& E_{\sigma\sigma'} e^{\im \sigma \varphi}  \left( \bm{\hat{e}}_{\sigma'} +\im \sigma \frac{\sqrt{2}}{kr}\frac{\sigma\cos\theta - \sigma'}{\sin\theta} \e_r + ... \right) ,\label{Fields}  \ee
where the last identity corresponds to the medium-far field expansion with
  \be
\frac{E_{\sigma \sigma'}}{E_0} &=& 
 \frac{e^{ikr}}{4\pi kr}  k^3 \left( \frac{\sigma\alpha_{\rm{E}} + \sigma' \alpha_{\rm{M}}}{2} \right) (\sigma\cos\theta + \sigma' )  , \\ \bm{\hat{e}}_{\sigma'} &=& \frac{1}{\sqrt{2}}( \bm{\hat{e}}_\theta + \im \sigma' \bm{\hat{e}}_\varphi) . \ee
The scattered fields by HRI dielectric nanoparticles  present  a number of peculiar properties arising from the interference between the electric and magnetic dipolar radiation and have been largely discussed both theoretical and experimentally 
\cite{evlyukhin2010optical,garcia2011strong,geffrin2012magnetic,person2013demonstration,fu2013directional,shi2013monodisperse,kuznetsov2016optically}.
Most of these properties are encoded in the far-field radiation pattern, i.e. in  the differential scattering cross section   given by \cite{nieto2011angle}
\be \label{S:3}
 \frac{{\rm{d}} \sigma_{\text{scat}} (\theta)}{\rm{d} \Omega}  &=&
 \lim_{r\rightarrow \infty} r^2 \frac{{\bf{S}}^{\text{scat}}\cdot  \e_r }{\left|{\bf{S}}^{(0)}\right|} = r^2 \frac{ |E_{\sigma+}|^2 +|E_{\sigma-}|^2}{|E_0|^2} \nonumber \\ &=&
 \frac{k^4 |\alpha_{\rm{sum}}|^2}{(4 \pi) ^2 }
\left( \frac{1+\cos^2 \theta}{2}+2g \cos \theta \right) ,  \ee
where ${\bf{S}}^{\text{scat}} = (1/2) \text{Re}\left\{{{\bf{E}}^{\text{scat}}}^* \times {\bf{H}}^{\text{scat}} \right\}$ is the time averaged Poynting vector, $ | \alpha_{\rm{sum}}|^2 \equiv  |\alpha_{\rm{E}}|^2+|\alpha_{\rm{M}}|^2$ and
\be
g =\frac{\text{Re}\left\{ \alpha_{\rm{E}} \alpha^*_{\rm{M}} \right\}}{|\alpha_{\rm{sum}}|^2} 
\ee
is the so-called asymmetry factor \cite{bohren2008absorption} for dipolar electric and magnetic scatterers \cite{nieto2011angle,gomez2012negative}.  

Although in the strict far field limit the flow lines of ${\bf{S}}^{\text{scat}} $ lie along the spherical radial direction, tracing them to their source, they do indeed spiral towards the origin in analogy with the light scattered by an electric dipole excited by circularly polarized light \cite{gough1986angular,arnoldus2004dipole,schwartz2006backscattered,arnoldus2008subwavelength,haefner2009spin}. Consequently, as sketched in Fig.  
\ref{delta}, the full Poynting vector ${\bf{S}}^{\text{scat}}$ makes an angle with the line of sight, which determines an apparent shift $\bm \Delta$ in the perceived position of the particle, with 
\be
\bm{\Delta}  &=& \lim_{kr \rightarrow \infty} -r \left( \frac{  {\bf{S}}^{\text{scat}} - \e_r \left(\e_r\cdot {\bf{S}}^{\text{scat}}\right)}{|{\bf{S}}_r|}\right)  \\ &=&
\lim_{kr \rightarrow \infty}  \left( \frac{  \e_r \times  \left(\r \times \bf{S}^{\text{scat}}\right)}{|{\bf{S}}_r|}\right) 
\\ &=& 
\lim_{kr \rightarrow \infty}  \left(   \frac{ 2\im}{k \left|\E^{\text{scat}}_\sigma\right|^2} \frac{ {\E_\sigma^{\text{scat}}}^* }{\sin\theta} \cdot\frac{\partial \E_\sigma^{\text{scat}}}{\partial \varphi}   \right) \e_\varphi 
\label{defDelta} 
\ee
where $\E_\sigma^{\text{scat}}$  is given by  Eqs. \eqref{Eff} and \eqref{Fields}. 
%\be
%{\bf{j}}_\infty =   \im \frac{ {\E_\sigma^{\text{scat}}}^* }{\sin\theta} \cdot\frac{\partial \E_\sigma^{\text{scat}}}{\partial \varphi}   \e_\theta =  - \frac{{\E_\sigma^{\text{scat}}}^* {\rm{L}}_z \E_\sigma^{\text{scat}}}{\sin\theta}   \e_\theta 
%\ee
Taking into account Eq. \eqref{OAMpf},  the apparent shift can be written as 
\be
\frac{\bm{\Delta}}{(\lambda / \pi)}   &=& - \frac{\ell_{\rm{z}}(\theta)}{\sin \theta} \bm{\hat{e}}_{\varphi} =
\frac{ s_z(\theta) - \sigma}{\sin \theta} \bm{\hat{e}}_{\varphi} 
\\ &=& -\sigma  \left[ \frac{  \sin \theta \left(1+ 2g \cos \theta  \right)}{  1+\cos^2 \theta   +4  {g} \cos \theta } \right]  \bm{\hat{e}}_{\varphi } . \label{S:5}
\ee
This is the first important result of this Letter: the shift is always along $\e_\varphi$, perpendicular to the incidence plane and proportional to the $z$-component of the OAM per photon. Importantly,   the sign of the displacement is purely determined by the incoming helicity.
% In particular, the azimuthal component of the Poynting vector takes the following form
%\be
%\bm{S}_{\varphi} \sim \frac{1}{Z} \frac{ |E_{\sigma+}|^2(\sigma -\cos\theta) + |E_{\sigma-}|^2(\sigma +\cos\theta)}{kr\sin\theta} \bm{\hat{e}}_\varphi. 
%\ee

\begin{figure}
\captionsetup{justification= raggedright}
\includegraphics[width=1\columnwidth]{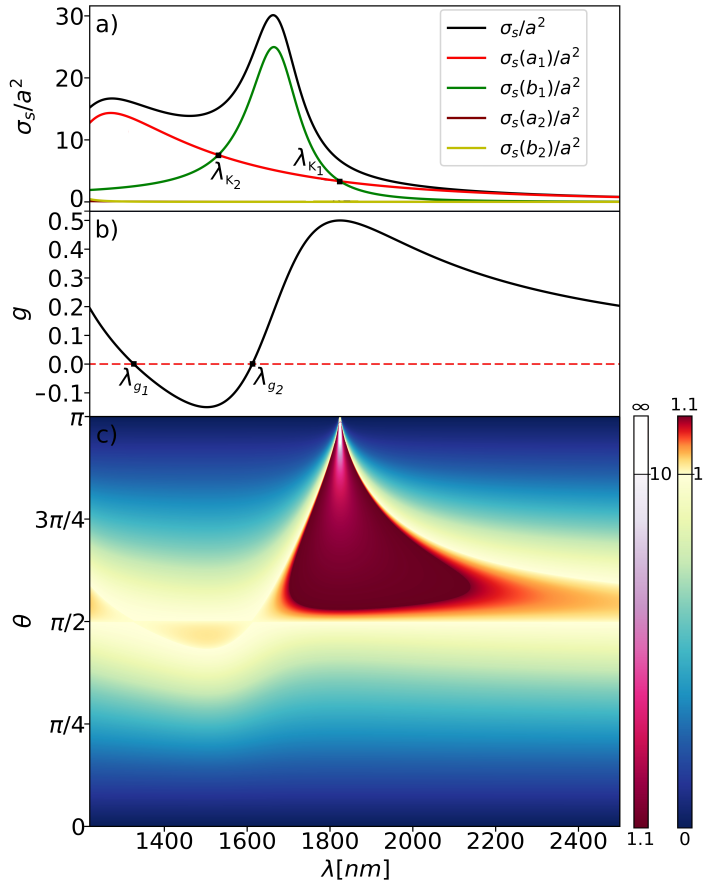}
\caption{(a) Scattering cross sections $\sigma_{\rm{s}}$ for a 230 nm Si nanosphere versus the wavelength. The special wavelengths $\lambda_{\rm{K_1}} = 1825$ nm and $\lambda_{\rm{K_2}} = 1520$ nm correspond to first and second Kerker conditions, respectively. (b) Asymmetry factor  versus the wavelength. This is identical to zero at $\lambda_{\rm{g_1}} = 1326$ nm and $\lambda_{\rm{g_2}} = 1612$ nm (and negative in between). The maximum value is localized at the first Kerker condition, namely, $\lambda_{\rm{K_1}} $. (c) Colormap of the   normalized optical mirage, $\Delta/(\lambda/\pi)$, versus the scattering angle and the wavelength. The maximum enhancement for $\lambda_{\rm{K_1}}$ at backscattering ($\theta = \pi$) is clearly observed. }\label{colores} 
\end{figure}

 In absence of magnetic dipolar response, setting $g$=0 in Eq. \eqref{S:5}, one recovers the previously reported results for electric dipoles \cite{arnoldus2008subwavelength,haefner2009spin}, which were interpreted as a result of transfer from SAM to OAM \cite{haefner2009spin,bliokh2011spin}. According to those previous works, this transfer is expected to be maximum at those directions at which the scattered light is linearly polarized (being the SAM of scattered photons identically zero). For an electric dipole excited by circularly polarized light the maximum transfer would take place in the plane perpendicular to the incoming light ($\theta = \pi/2$) being the maximum displacemente equal to $\Delta = \lambda/\pi$.

%\AGE{However, the fields scattered by an electric and a magnetic dipole are generally elliptically polarized \cite{garciaetxarri2017,garciaetxarri2013}.} In particular, excited by circularly polarized light, the light on the plane perpendicular to the incoming light is no longer linearly polarized. Nonetheless, it is worth noticing that the presence of an additional magnetic contribution does not modify the streamlines of the Poynting vector on the $\theta=\pi/2$ plane (as shown in Fig. \ref{flechas}) leading to the same subwavelength displacement. \AGE{However, in contrast with the electric dipolar case,} out of this plane, the apparent displacement presents a peculiar behaviour that strongly depends on both $\theta$ and wavelength.

The fields scattered by  electric and magnetic dipoles present a very different polarization structure \cite{garciaetxarri2017,garciaetxarri2013}. Contrary to the purely electric (or magnetic) case, when excited with a circularly polarized field, the scattered radiation on the plane perpendicular to the incoming light ($\theta=\pi/2$) is no longer linearly polarized. Interestingly, this change does not affect the streamlines of the Poynting vector on this particular plane (as shown in Fig. \ref{flechas}), leading to the same subwavelength optical mirage. However, out of this plane the apparent displacement presents a peculiar behaviour that strongly depends on both $\theta$ and the wavelength.

\begin{figure}
\includegraphics[width=1\columnwidth]{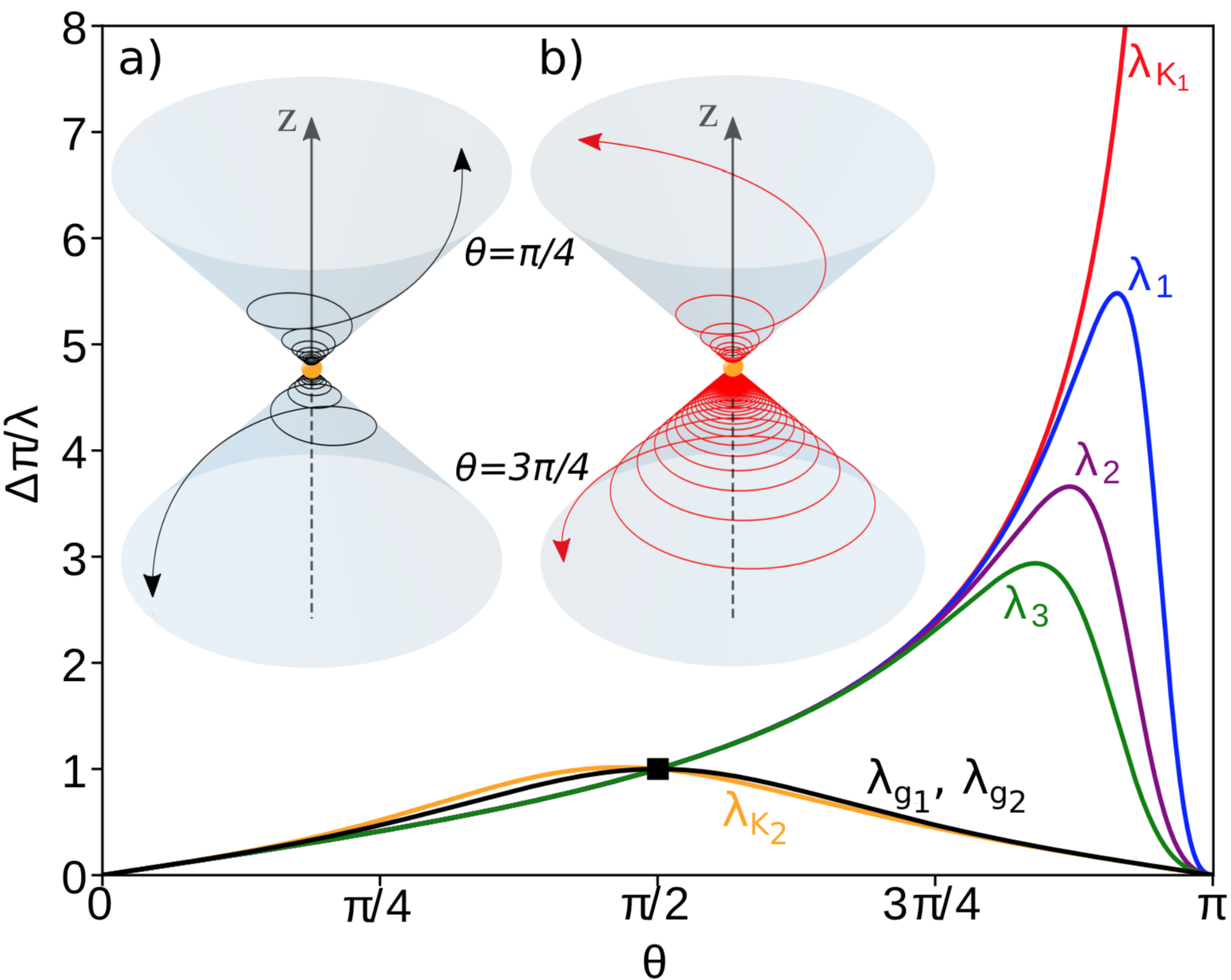}
\captionsetup{justification= raggedright}
\caption{Optical mirage colormap (Fig. \ref{colores}) cuts versus the scattering angle for different values of the wavelength, belonging to regions with $g < 0$ ($\lambda_{\rm{K_2}}$), $g = 0$ ($\lambda_{\rm{g_1}}, \lambda_{\rm{g_2}}$) and $g > 0$ ($\lambda_1, \lambda_2$ and $\lambda_3$, respectively decreased 5, 10 and 15 nm with respect to $\lambda_{\rm{K_1}}$, and $\lambda_{\rm{K_1}}$ itself). At $\theta = \pi /2$,  $\Delta = \lambda / \pi$ is observed to be a universal value . Both subplots show examples of trajectories of the Poynting vector at forward and backscattering, being similar for $\lambda_{\rm{g_1}}$ and $\lambda_{\rm{g_2}}$ (a) and considerably different for $\lambda_{\rm{K_1}}$ (b).}\label{cut} 
\end{figure}

Figures \ref{colores} and \ref{cut} summarize the anomalous behavior of the apparent displacement $\Delta(\lambda,\theta)$ for silicon nanospheres in the infrared (similar behavior is obtained in other spectral ranges as long as the scattering cross section can be described by only the first two dipolar multipoles, see Fig. \ref{colores}a). As it can be seen in Figure \ref{cut}, for $\theta=\pi/2$ the displacement is always $\lambda/\pi$ for all wavelengths. When the asymmetry factor $g$ is negative ($\lambda_{g1}< \lambda < \lambda_{g2}$) the maximum displacement occurs for $\theta < \pi/2$ and it is always subwavelength but slightly larger than the one for $\theta=\pi/2$. However, for $g>0$  the apparent displacement can be much  larger than $\lambda/\pi$ and when the electric and magnetic polarizabilities are identical ($\lambda = \lambda_{\rm{K}1}$), i.e. at the so-called first Kerker condition, it diverges as $\theta \rightarrow \pi$. Notice that the singularity is resolved naturally since at the first Kerker condition there is exactly zero back-scattered intensity. 

We can now examine the peculiar behaviour of $\bm{\Delta}$ near the first Kerker condition in terms of the angular momentum flow. When the electric and magnetic responses are identical, i.e. $\alpha_{\rm{E}} = \alpha_{\rm{M}}$, the system is ``dual'' and the scattering preserves helicity \cite{fernandez2013electromagnetic,schmidt2015isotropically}. In this case, the asymmetry factor is maximum, $g = 1/2$, (see Fig \ref{colores}b) which leads to
% \be
% \ell(\theta) = \sigma (1- \cos \theta) \qquad s_z(\theta) = \sigma \cos\theta
% \ee
$
s_z(\theta) = \sigma \cos\theta
$ 
and
\begin{equation}\label{S:8}
\frac{\bm{\Delta} \pi }{\lambda_{\rm{K_1}}} = - \sigma \tan \left( \frac{\theta}{2} \right)  \bm{\hat{e}}_{\varphi}.
\end{equation}
From this equation two interesting limiting cases can be identified: %The forward and  perpendicular direction, i.e. $\theta =0$ and $\theta = \pi/2$, respectively. 
firstly, in the forward direction the optical mirage and $l_{\rm{z}}$ go to zero since ${\bf{S}}_{{\varphi}} = 0$. This can alternatively be understood by means of the symmetries of the system: being the scatterer dual, the system must conserve the helicity of the incoming field, which in the forward direction corresponds to the spin density. Thus, the incident circular polarization is preserved in the forward direction and must carry all the angular momentum density (leaving $\ell_{\rm{z}}$=0). Secondly, in the direction perpendicular to the incident wave-vector ($\theta=\pi/2$), %it is clear that the electric and magnetic dipoles do not \AGE{interfere} since 
the interference term vanishes. As a consequence, $s_z=0$ and $\ell_z=\sigma$ and, in analogy with electric dipoles, we obtain ${\Delta} =\sigma  \lambda/\pi $, although in that case light in this direction is fully circularly polarized  (see Fig. \ref{flechas}). 

The most striking effect arises at an observation angle near backscattering $\theta \lesssim \pi$ where, as discussed above, the apparent displacement diverges. This divergence is solved because the Poynting vector becomes strictly zero at backscattering, which suggests the appearance of an optical vortex in that direction. As a matter of fact, near backscattering $\ell_z(\lesssim \theta) \rightarrow 2\sigma$, while the spin reverses sign $s_z(\theta\lesssim \pi) \rightarrow -\sigma$
 (but still maintaining constant helicity), which confirms the existence of a vortex with $l=2\sigma$ emerging from a nanoparticle as a nanoscale analogue of the light backscatterd from a perfect reflecting cone 
\cite{mansuripur2011spin}. 

In conclusion, we have shown that light scattering from dipolar electric and magnetic nanoparticles, excited by circular polarized light, can lead to optical mirages values much larger than the incident wavelength. The properties of the optical mirage were discussed in terms of spin-orbit interactions and helicity conservation. Interestingly, for dual spheres, i.e. at the so-called first Kerker condition, we predicted  a huge enhancement of the apparent shift  related to the emergence of an optical vortex in the backscattering direction. 

This research was supported by the Spanish Ministerio de Econom\'{\i}a y Competitividad (MICINN) and European Regional Development Fund (ERDF)  Projects  FIS2014-55987-P,  FIS2015-69295-C3-3-P and FIS2017-82804-P, by the Basque 
Dep. de Educaci\'on  Project PI-2016-1-0041 and by Basque Government ELKARTEK program through MICRO4FAB (KK-2016/00030) and $\mu 4 F$ (KK-2017/00089) Projects.  A. G.-E. received funding from the Fellows Gipuzkoa fellowship of the Gipuzkoako Foru Aldundia through FEDER "Una Manera de hacer Europa"  . 

Jorge Olmos-Trigo and Cristina Sanz-Fern\'andez contributed equally to this work.

\bibliography{Poy_bib}

\end{document}